\documentclass[%
 aip,
 amsmath,amssymb,
 reprint,%
]{revtex4-1}

\usepackage{graphicx}
\usepackage{dcolumn}
\usepackage{bm}

\usepackage[utf8]{inputenc}
\usepackage[T1]{fontenc}
\usepackage{mathptmx}
\usepackage{etoolbox}

\usepackage[dvipsnames]{xcolor}
\usepackage{makecell}

\usepackage{tabularray}
\UseTblrLibrary{diagbox} 

\makeatletter
\def\@email#1#2{%
 \endgroup
 \patchcmd{\titleblock@produce}
  {\frontmatter@RRAPformat}
  {\frontmatter@RRAPformat{\produce@RRAP{*#1\href{mailto:#2}{#2}}}\frontmatter@RRAPformat}
  {}{}
}%
\makeatother
\begin{document}

\preprint{AIP/123-QED}

\title[Sample title]{Birefringent-colored optical profiling of wood presenting the 3D cellulose microfibril architecture\vspace{0.5cm}}

\author{Jieh-Wen Tsung}
 \email{Contact author: jiehwen.tsung@nycu.edu.tw}
\affiliation{%
 Department of Electrophysics, National Yang Ming Chiao Tung University, Hsinchu, 300110, Taiwan\\
}%
\affiliation{Center for Emergent Functional Matter Science, National Yang Ming Chiao Tung University, Hsinchu 300093, Taiwan}

\date{\today}

\begin{abstract}
Wood slices under a polarized optical microscope show a spectrum of interference colors, such as cyan, blue, magenta, yellow, and bluish gray because of the birefringent cellulose in the microfibrils.
An optical model is established to simulate the birefringent-colored micrograph of wood. Five typical cell wall architectures, line, helix, ring, crossed helix, and twisted helix, are considered. With a retardation wave plate to distinguish fibrils of different orientations, each structure displays its unique birefringent-colored texture, presenting its underlying 3D structure with the vivid colors.
Cross sections of the trunk and twig of \textit{Eucalyptus grandis} presented birefringent-colored profiles.
Three identification methods were compared: visible fibril trends, birefringent-colored optical textures, and the simulated look-up library. The three methods gave consistent results, proving that the birefringent-color tags are efficient and accurate.
Electron and atomic force microscopy are unable to resolve cellulose microfibrils embedded in the lignin and hemicellulose matrix. Polarized optical microscopy overcomes this by selectively detecting birefringent cellulose, enabling accurate identification of complex helical structures.
This method enables statistical and spatial analysis of complex biomaterial compositions.
Capable of profiling dozens of cells simultaneously, this high-throughput optical method provides a potentially fully automated analysis framework for plant science, biomechanics, and bioinspired cellulose materials.
\end{abstract}

\maketitle

%

\section{\label{sec:level1}Introduction}


Wood is a composition of slender, long, tube-like cells. The cell walls are composed of cellulose fibrils embedded in a lignin and hemicellulose matrix.
The 3D fibril organization explains how wood achieves extraordinary tensile strength, structural integrity, flexibility, and energy-saving water transport~\cite{MFA_review_Barnett, BioMaterial_why_strong}.
Therefore, the 3D structure of the fibrils in the cell wall has been of great interest in biophysics and biomimetics.


Five typical microfibril organizations have been found in the plant cell walls~\cite{Xylem_5helix}:
line, 
helix, 
ring, 
crossed helix, 
and twisted helix (helicoid)~\cite{CLC_fibril_Bouligand, CLC_vivo_Bouligand}, see Table~\ref{table:intro_5helix}.
Fibrils arrange longitudinally (line), helically (helix), or azimuthally (ring).
A crossed helical cell wall has alternating left- and right-handed layers.
In a twisted helix, the microfibrils rotate progressively across the thickness of the cell wall.
Xylem cells in trees (wood) and rigid cells in grass, which develop secondary cell walls, contain the S1, S2, and S3 sublayers~\cite{MFA_review_Barnett}. The helical structures are significant in the thickest S2 layer, while the primary wall, S1, and S3 are thin layers of crossed or aligned fibrils.

\begin{table*}
\caption{\label{table:intro_5helix}Cellulose microfibril structure of xylem cell walls and examples.}
\begin{ruledtabular}
\begin{tabular}{cccccc}
\makecell[c]{}
&\begin{minipage}{0.12\textwidth}
      \includegraphics[width=\linewidth]{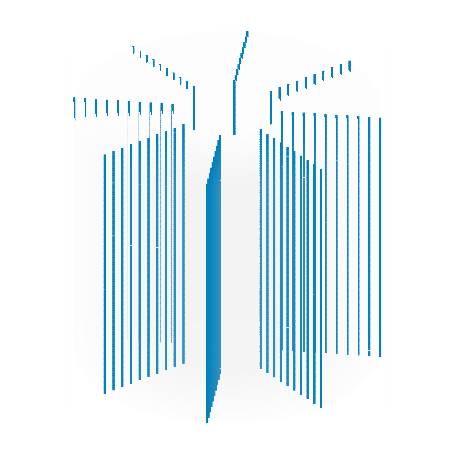}
    \end{minipage}
&\begin{minipage}{0.12\textwidth}
      \includegraphics[width=\linewidth]{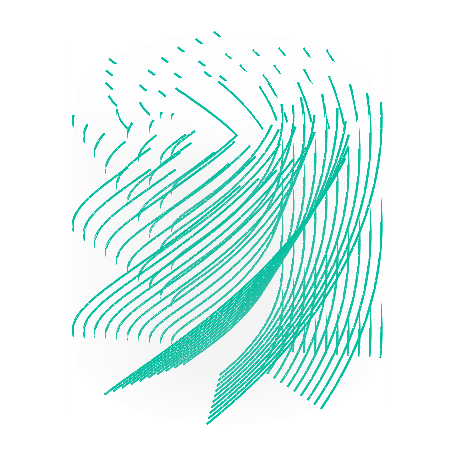}
    \end{minipage}
&\begin{minipage}{0.12\textwidth}
      \includegraphics[width=\linewidth]{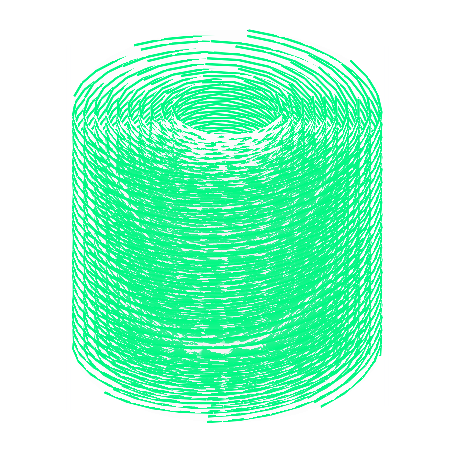}
    \end{minipage}
&\begin{minipage}{0.12\textwidth}
      \includegraphics[width=\linewidth]{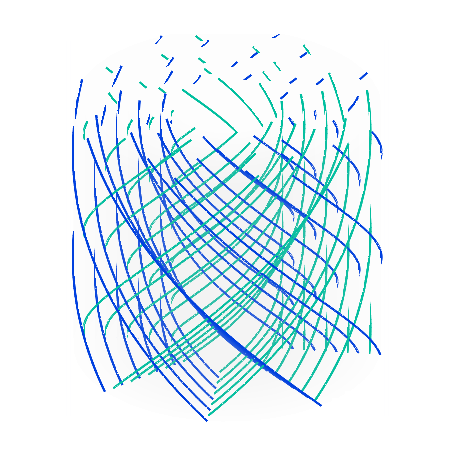}
    \end{minipage}
&\begin{minipage}{0.12\textwidth}
      \includegraphics[width=\linewidth]{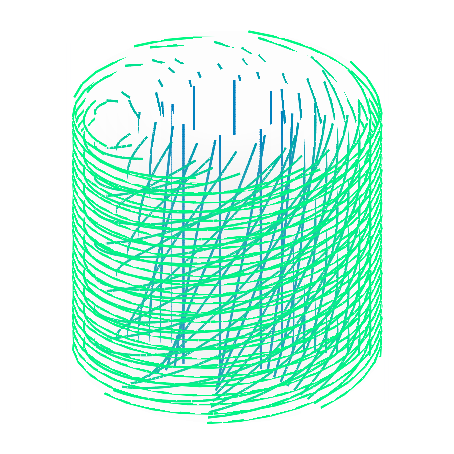}
    \end{minipage} \\
\textbf{}
& \textbf{Line}
& \textbf{Helix}
& \textbf{Ring}
& \textbf{Crossed helix}
& \textbf{Twisted helix}\footnote{Also called helicoid or Bouligand structure.}\\
\hline
Examples
& \makecell[c]{Tension wood\cite{MFA_review_Barnett, Helix_CrossHelix_Eder}}
& \makecell[c]{S2 of \\ xylem cells\cite{MFA_review_Barnett, Helix_CrossHelix_Eder}}
& \makecell[c]{S1 S3 of \\ xylem cells\cite{MFA_review_Barnett, Helix_CrossHelix_Eder}}
& \makecell[c]{Grass\cite{CrossedHelix_Bamboo} \\ Conifer\cite{CLC_nascent_Reis}}
& \makecell[c]{Stone cell\cite{Plant_cell_LC_Neville_1993} \\ Seed coat\cite{twistedHelix_seed_shell} \\ fruit peel\cite{CLC_fruit_Vignolini}}\\
\hline
\makecell[c]{Where to find it \\ in our eucalyptus tree}
& \makecell[c]{Bark of the twig}
& \makecell[c]{In the twig}
& \makecell[c]{Vessels in the twig \\ and the trunk}
& \makecell[c]{In the twig}
& \makecell[c]{In the trunk}\\
\end{tabular}
\end{ruledtabular}
\end{table*}


Scanning Electron Microscope (SEM), Transmission Electron Microscope (TEM), Atomic Force Microscope (AFM), and Confocal Laser Scanning Microscope (CLSM) struggle with wood, mainly because the cellulose fibrils are embedded in the lignin and hemicellulose matrix. The dehydration, sputter-coating, or cryo-fixation could destroy the helical microstructure, too. Although the resolution is sufficient for showing fibrils~\cite{Microscopy_for_Wood}, the actual image is often limited to recognizing the S1, S2, and S3 layers~\cite{SEM_fiber_vessel_ray}. Identifying twisting fibrils is very difficult.
To overcome the complications and damage, we apply polarized optical microscopy (POM), since the cellulose fibrils are birefringent~\cite{CLC_living_Mitov, CLC_helical_Godinho, CLC_living_Godinho} but the matrix is optically isotropic.

\begin{figure*}
\centering
\includegraphics[width=1.0\textwidth]{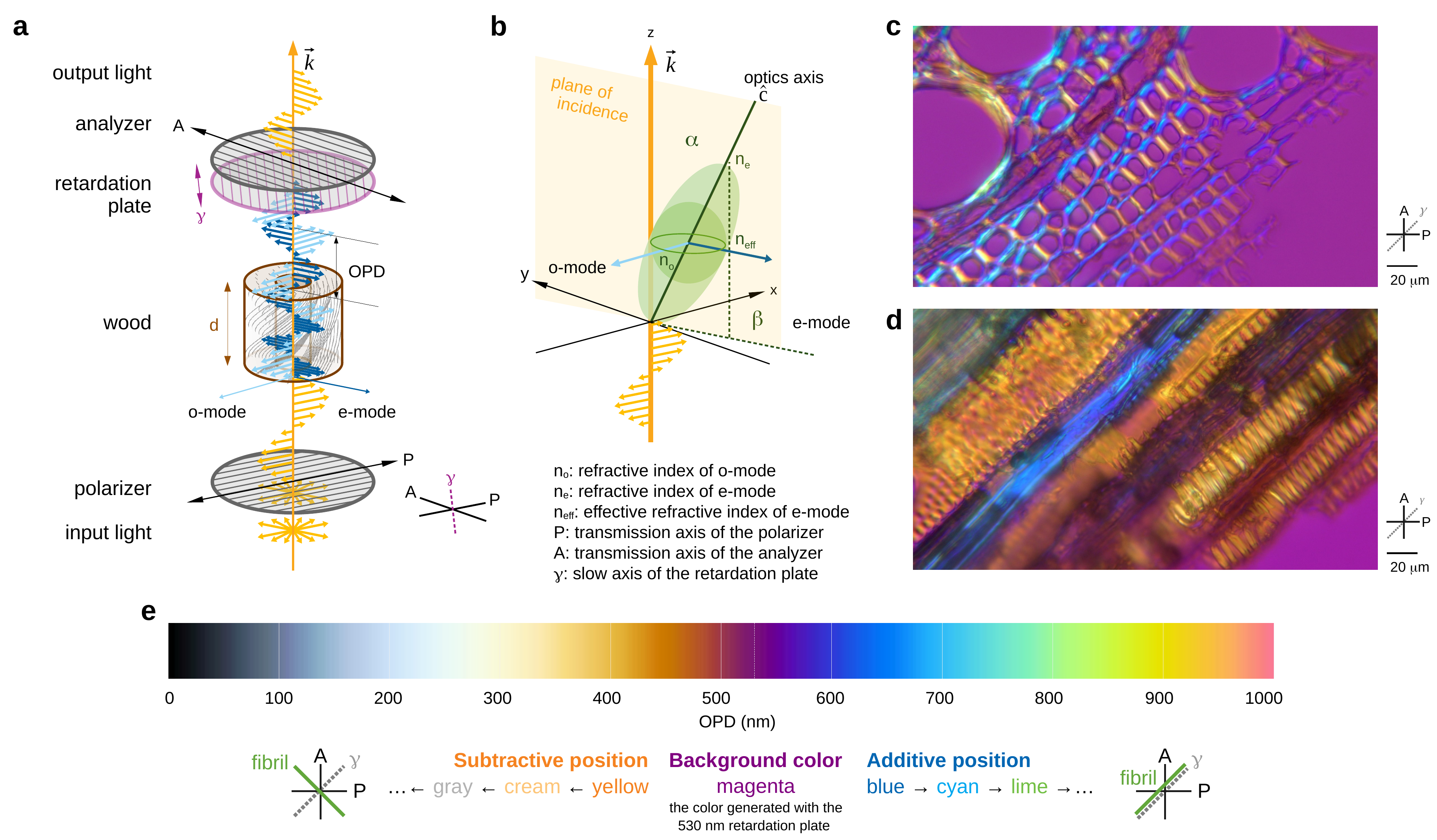}
\caption{Polarized optical microscopy for wood.
(a) Schematic diagram of light propagation ($\vec k$) through birefringent cellulose fibrils in wood, decomposing into ordinary (o-mode) and extraordinary (e-mode) modes with optical path difference $\text{OPD}$.
(b) Refractive index ellipsoid of the cellulose fibril.
(c-d) POM images of transverse and longitudinal cross sections.
(e) Birefringent colors versus $\text{OPD}$ to identify the fibril direction.}
\label{fig:POM}
\end{figure*}

Wood slices under POM are colorful. Beyond the beauty, the colors indicate the orientation of the cellulose fibrils~\cite{POM_Bloss, POM_Delly}.
Light passing through the birefringent material decomposes into ordinary (o-mode) and extraordinary (e-mode) modes that propagate at different speeds (Fig.~\ref{fig:POM}(a) and (b)).
The phase shift renders vivid birefringence colors~\cite{Birefringence_color_chart_Sorensen, Birefringence_color_chart_Johnsen} (Fig.~\ref{fig:POM}(c) and (d)).
With polarizer, analyzer, and retardation plates,
the interference color and the brightness can be quantitatively translated into the polar ($\alpha$) and azimuthal angles ($\beta$) of the microfibril orientation, respectively.
Therefore, without breaking, freezing, drying, or coating the plant tissue, the orientation of the fibrils can be clearly displayed as the colors in the microscopic image.


POM has been applied to measure the microfibril angles of wheat cells~\cite{MFA_POM_SEM}, but the identification of 3D structure has never been successful.
The major challenge is the optical degeneracy, since the fibrils on the diagonal directions ($45^\circ$ and $135^\circ$) show identical signals in crossed polarizer ($0^\circ$) and analyzer ($90^\circ$).
Therefore, the optical profile of the helix, crossed helix, and twisted helix can be difficult to distinguish.
This research provides a major improvement by building the optical model for the 3D helices and introducing the retardation wave plate to distinguish fibrils oriented in different directions.

The optical model of the line, helix, ring, crossed helix, and twisted helix of cellulose fibrils is established, and the birefringent-colored optical profiles of the transverse and longitudinal cross sections are simulated to prepare a look-up library.
The results confirm that the five typical organizations have their unique birefringent colors, enabling the accurate identification.
Slices of \textit{Eucalyptus grandis} were examined under the microscope.
We selected five xylem cells known to have line, helix, ring, crossed helix, and twisted helix structures, and analyzed the color distribution in the POM images. The actual color distributions were consistent with the simulated results, demonstrating that optical profiling is effective and reliable.

\section{\label{sec:model}Optical Modeling}

\subsection{Propagation of light in birefringence material}

As light travels through a slab of birefringent material, the optical path difference ($\text{OPD}$) between extraordinary (e-mode) and ordinary (o-mode) modes is
\begin{equation} \label{eq:OPD}
\text{OPD} = (n_{eff} - n_o) \cdot d
\end{equation}
where
\begin{equation} \label{eq:n_eff}
n_{eff} = \frac{n_o n_e}{\sqrt{(n_e\cos\alpha)^2 + (n_o\sin\alpha)^2}}
\end{equation}
where $d$ = sample thickness; $n_e$, $n_o$ = extraordinary (e-mode) and ordinary (o-mode)
refractive indices; $n_{eff}$ = effective e-mode index; $\alpha$ = angle between 
light propagation ($\hat k$) and fibril optics axis ($\hat c$). The angle $\alpha$ 
determines birefringence ($\Delta n = n_{eff}-n_o$) and thus OPD.

The change of the polarization state can be calculated with the Jones matrix~\cite{Optics_Hecht, Optics_LC_Yeh}
\begin{align*}
\mathbf{J} &=\left[ {\begin{array}{cc}
    \cos{\beta}  &-\sin{\beta} \\
    \sin{\beta}  &\cos{\beta} \\
    \end{array} } \right]
    \left[ {\begin{array}{cc}
    e^{-i\frac{\delta}{2}} & 0 \\
    0                      & e^{+i\frac{\delta}{2}} \\
    \end{array} } \right]
    \left[ {\begin{array}{cc}
    \cos{\beta}  &\sin{\beta} \\
    -\sin{\beta}  &\cos{\beta} \\
    \end{array} } \right] \\
\end{align*}
where $\delta$ is optical phase retardation, $2 \pi \frac{\text{OPD}}{\lambda}$; $\beta$ is the angle between the projection of $\hat c$ on the polarization plane of the light and the reference axis (x axis).
If $\alpha$ and $\beta$ are constants in the entire slab, the OPD is simply $\Delta n \cdot d$.

The polarization state of the output light is 
\begin{equation}
\vec{E}_{o} = \mathbf{J} \times \vec{E}_{i}
\end{equation}
In crossed polarizer and analyzer, the intensity of the output light is
\begin{equation} \label{eq:I_bi_color}
\frac{I}{I_0} = \sin ^2{ \left( 2\beta \right) } \sin^2{  \left(\pi\frac{\text{OPD}}{\lambda} \right) }
\end{equation}
The color of the transmitted light (determined by wavelength $\lambda$) is related to $\text{OPD}$, which is determined by the polar angle $\alpha$. The brightness indicates the azimuthal orientation $\beta$.
Birefringence color as a function of $\text{OPD}$ is shown in Fig.~\ref{fig:POM}(e).

\subsection{Vectorial summation}



For progressive axis variation, the sample was modeled as thin layers:
\begin{equation}
\vec E_{o} = \mathbf{J}_n \times \mathbf{J}_{n-1} \times \cdot \cdot \cdot \times \mathbf{J}_{2} \times \mathbf{J}_{1} \times \vec E_{i}
\end{equation}
The matrix multiplication must be in sequence, since the Jones matrices do not commute.
To reduce computation, infinitesimal thin-layer approximation was applied, allowing matrix multiplication to be simplified to vectorial summation~\cite{Vectorial_Sum_Goldstein} of orthogonal phase components in the phase-retardation vector space:
\begin{equation}
\begin{aligned}
\vec{\Gamma}_{s}
& = \left[ {\begin{array}{cc}
X \\
Y
\end{array} } \right]
= \int_{0}^{d} \frac{2\pi}{\lambda}(n_{eff}(z) - n_o)
\left[ {\begin{array}{cc}
\cos(2\beta(z)) \\
\sin(2\beta(z))
\end{array} } \right]
dz \\
& = \begin{bmatrix}
    \text{component in the plane of incidence} \\
    \text{component normal to the plane of incidence} \\
\end{bmatrix}
\end{aligned}
\end{equation}
where the light propagates ($\vec k$) on the $z$ axis from the entry face ($z=0$) to the exit face ($z=d$). The incident light is decomposed into two components that are in and normal to the plane of incidence. The phase accumulation of each component is tracked independently. Finally, the two components recombine, and the total phase retardation ($\delta_{s}$) and the effective optics axis orientation ($\beta_{s}$) are calculated as
\begin{align*}
\delta_{\text{s}} = \sqrt{X^2 + Y^2}
\end{align*}
\begin{align*}
\beta_{\text{s}} = \frac{1}{2} \tan^{-1}\left(\frac{Y}{X}\right)
\end{align*}
When the sample is placed between crossed polarizer and analyzer, the transmittance becomes
\begin{equation} \label{eq:I_bi_color_vec}
\frac{I}{I_0} = \sin ^2{ \left( 2\beta_{s} \right) } \sin^2{  \left(\frac{\delta_{s}}{2} \right) }
\end{equation}

\subsection{Colorizing the fibril direction}

Using crossed polarizer and analyzer, 
fibrils with $\hat c$ orientated at $45^\circ$ and $135^\circ$ show the same color.
To distinguish the two cases, a wave plate with retardation of $\delta_r$ and the slow axis at $\beta_\gamma$ is placed above the sample. The total optical retardation becomes
\begin{equation}
\begin{aligned}
\vec {{\Gamma}}_{total} &= \vec{\Gamma}_{\gamma} + \vec{\Gamma}_{s} \\
&= \delta_{\gamma} \begin{bmatrix} \cos (2\beta_{\gamma}) \\ \sin (2\beta_{\gamma}) \end{bmatrix} + \delta_{s} \begin{bmatrix} \cos (2\beta_{s}) \\ \sin (2\beta_{s}) \end{bmatrix} \\
\end{aligned}
\end{equation}
When $\beta_\gamma=\beta_s$ (\textbf{the additive position})
\begin{equation}
\begin{aligned}
\vec {{\Gamma}}_{total} &= ( \delta_{\gamma} + \delta_{s}) \begin{bmatrix} \cos (2\beta_{\gamma}) \\ \sin (2\beta_{\gamma}) \end{bmatrix}
\end{aligned}
\end{equation}
When $\beta_\gamma = \beta_s \pm 90^\circ$ (\textbf{the subtractive position})
\begin{equation}
\begin{aligned}
\vec {{\Gamma}}_{total} &= ( \delta_{\gamma} - \delta_{s}) \begin{bmatrix} \cos (2\beta_{\gamma}) \\ \sin (2\beta_{\gamma}) \end{bmatrix}
\end{aligned}
\end{equation}
Therefore, the beautiful colors in Fig.~\ref{fig:POM}(c) and (d) are indicators of the fibril directions.
The default setting of a POM is 530 nm of retardation with the slow axis $\gamma$ at $45^\circ$, generating the magenta background of the micrograph.
The fibrils on $45^\circ$ show blue, which is the birefringence color of (530 + 100) nm of OPD.
The fibrils on $135^\circ$ show yellow, which is the birefringence color of (530 - 100) nm of OPD.



\subsection{Quantitative look-up library}

A computational model was developed to simulate the birefringence colors of the line, helix, ring, crossed helical, and twisted helical cell walls under POM. GNU Octave was used for the simulations.

\subsubsection{Transverse cuts}

\begin{figure*}
\centering
\includegraphics[width=1.0\textwidth]{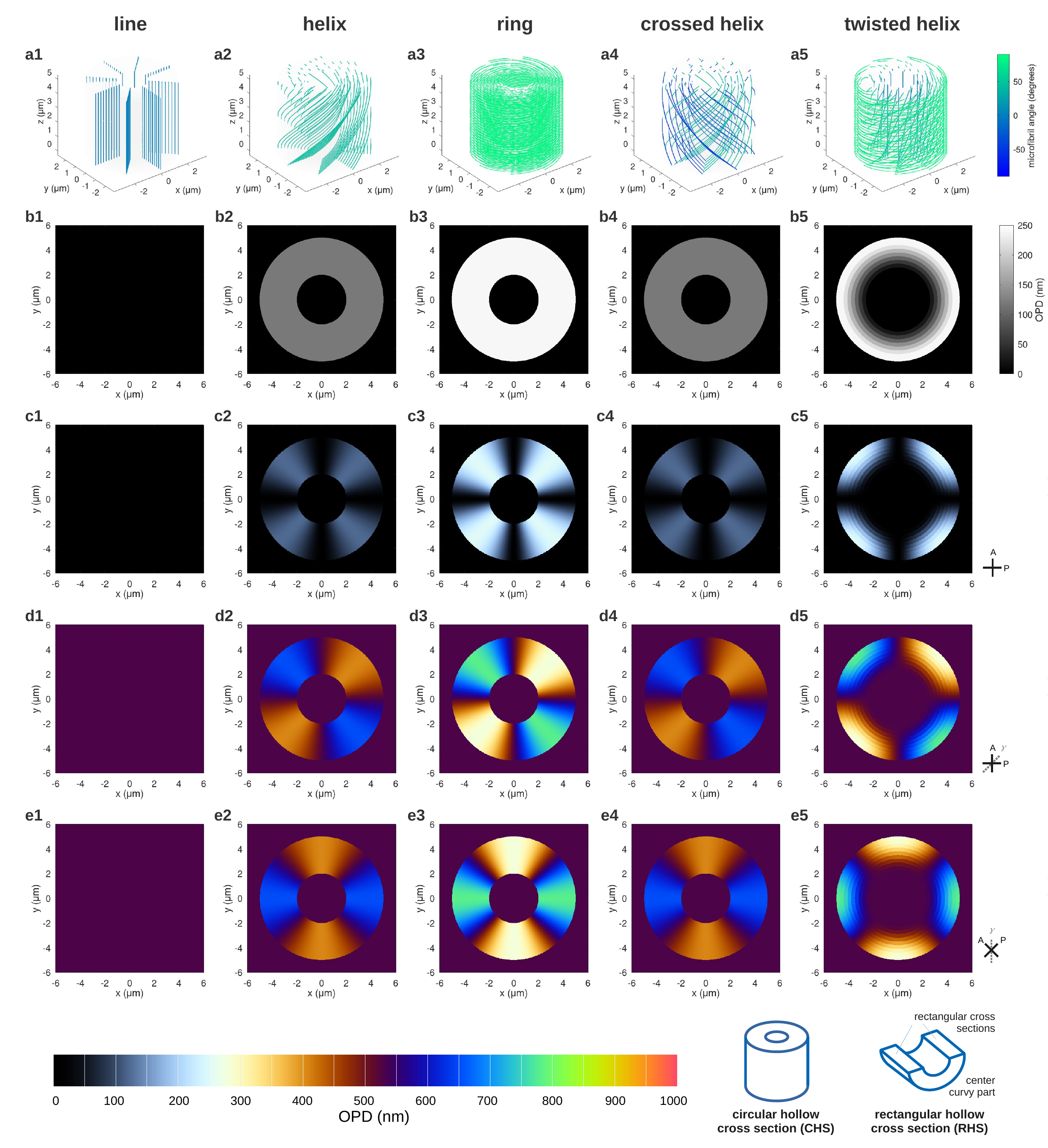}
\caption{Simulated POM images of the transverse cross sections.
(a1-5) 3D models with color-coded fibril angle.
(b1-5) Optical path difference (OPD) magnitude in gray scale.
POM images with (c1-5) P-A = $0^\circ$-$90^\circ$, (d1-5) P-A-$\gamma$ = $0^\circ$-$90^\circ$-$45^\circ$, and (e1-5) P-A-$\gamma$ = $45^\circ$-$135^\circ$-$90^\circ$.
P, polarizer.
A, analyzer.
$\gamma$, slow axis of the 530~nm retardation plate.
Simulation parameters: 
$n_o$, 1.53.
$n_e$, 1.58.
Thickness, 5~\textmu m.}
\label{fig:Xylem_MFA_5modes_opt_top}
\end{figure*}

A transverse cross section of a xylem cell was modeled as a hollow cylinder (Fig.~\ref{fig:Xylem_MFA_5modes_opt_top}(a1-5)), consisting of 10 layers of cylindrical shells. Each shell was a lamina of parallel cellulose fibrils with a specific $\alpha$. The fibrils had a right-handed twist~\cite{CLC_helical_Godinho, CLC_living_Godinho}.
The average refractive index of cellulose is 1.53, and the $\Delta n$ of wood pulp is in the range of 0.05$\pm$0.01~\cite{Delta_n_wood}.
For the simulation, $n_o$ and $n_e$ were taken as 1.53 and 1.58, respectively.
The height of the cylinder was 5~\textmu m. The inner and outer radius of the cylinder were 2~\textmu m and 5~\textmu m, respectively.

Light propagating along the cylinder axis ($\hat z$) showed $\text{OPD}$ increasing with $\alpha$.
Line, helix, and ring gave zero, moderate, and maximum OPD, respectively (Fig.~\ref{fig:Xylem_MFA_5modes_opt_top}(b1-3)).
Crossed helix with $\alpha$ on $+45^\circ$ and $-45^\circ$ showed identical $\text{OPD}$ (Fig.~\ref{fig:Xylem_MFA_5modes_opt_top}(b4)).
$\alpha$ of fibrils in twisted helix varied from $0^\circ$ (inner) to $90^\circ$ (outer), showing $\text{OPD}$ from 0 to the maximum magnitude (Fig.~\ref{fig:Xylem_MFA_5modes_opt_top}(b5)).

In crossed linear polarizer (P) and analyzer (A), the xylem slices showed birefringence colors from darkness to bluish gray (Fig.~\ref{fig:Xylem_MFA_5modes_opt_top}(c1-5)).
The fibrils with $\beta$ on $0^\circ$ and $90^\circ$ contributed zero OPD, so the vortex-shaped helical structures showed the four dark brushes of a Maltese cross~\cite{Textures_LC_Dierking}.
The line cell with all the fibrils vertically upright gave zero OPD and dark texture, too.

To distinguish the fibrils on $45^{\circ}$ and $135^{\circ}$, the 530 nm retardation plate with $\gamma$ on $45^{\circ}$ was placed between the sample and the A.
All the vortex-like helices showed the yellow of subtractive position at the 1st and 3rd quadrants and the blue of additive position at the 2nd and 4th quadrants (Fig.~\ref{fig:Xylem_MFA_5modes_opt_top}(d2-4)), except the line cell (Fig.~\ref{fig:Xylem_MFA_5modes_opt_top}(d1)) and the inner side of the twisted helix (Fig.~\ref{fig:Xylem_MFA_5modes_opt_top}(d5)), which showed magenta of vertical fibrils.

The vertical fibril, the fibril on $0^\circ$, and the fibril on $90^\circ$ showed the same magenta color.
When P-A-$\gamma$ rotated, vertical fibrils showed constant magenta color (Fig.~\ref{fig:Xylem_MFA_5modes_opt_top}(d1) and (e1)), and in-plane vortex fibrils displayed a rotating Maltese cross pattern.
In the following figures, we always show two typical cases:
P-A-$\gamma =$ $0^\circ$-$90^\circ$-$45^\circ$ (Fig.~\ref{fig:Xylem_MFA_5modes_opt_top}(d1-5))
and 
P-A-$\gamma =$ $45^\circ$-$135^\circ$-$90^\circ$ (Fig.~\ref{fig:Xylem_MFA_5modes_opt_top}(e1-5)).

The transverse sections of line and twisted helix had characteristic color distribution, so they could be easily identified. However, the helix, ring, and crossed helix looked very similar, especially when $\alpha$ was still unknown. Additional characteristics from the longitudinal cuts were required to ensure an accurate identification.

\subsubsection{Longitudinal cuts}

\begin{figure*}
\centering
\includegraphics[width=1.0\textwidth]{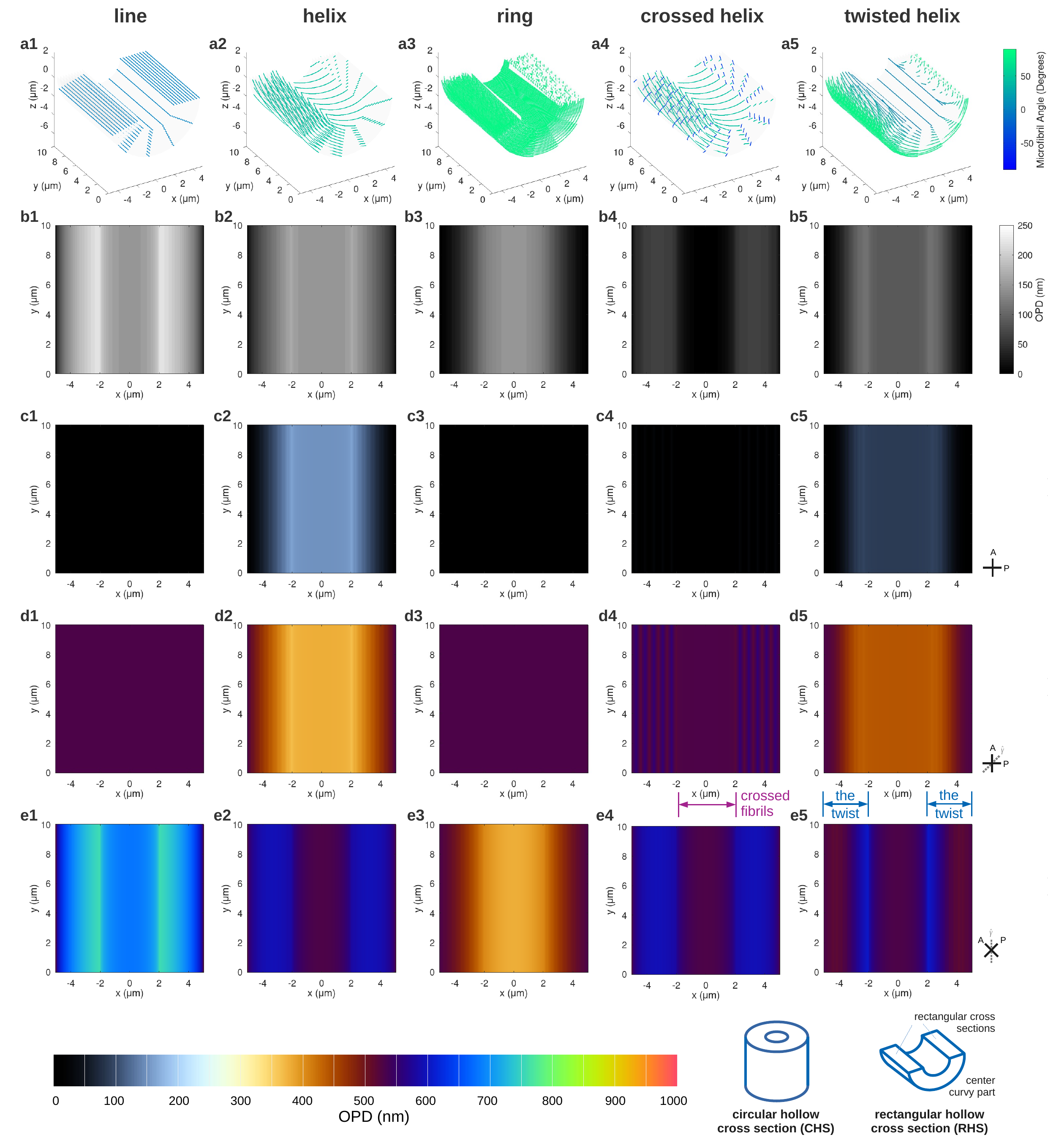}
\caption{Simulated POM images of the longitudinal cross sections.
(a1-5) 3D rectangular hollow cross sections with color-coded fibril angle.
(b1-5) Optical path difference (OPD) magnitude in gray scale.
POM images with (c1-5) P-A = $0^\circ$-$90^\circ$, (d1-5) P-A-$\gamma$ = $0^\circ$-$90^\circ$-$45^\circ$, and (e1-5) P-A-$\gamma$ = $45^\circ$-$135^\circ$-$90^\circ$.
P, polarizer.
A, analyzer.
$\gamma$, slow axis of the 530~nm retardation plate.
Simulation parameters: 
$n_o$, 1.53.
$n_e$, 1.58.
Inner radius, 2~\textmu m.
Outer radius, 5~\textmu m.
}
\label{fig:Fig_Xylem_MFA_5modes_opt_cut_half}
\end{figure*}

The longitudinal cross section of the cell wall was one-half of a hollow multi-layered cylinder (Fig.~\ref{fig:Fig_Xylem_MFA_5modes_opt_cut_half}(a1-5)).
The objective lens faced the rectangular hollow cross section (RHS) of the cylinder.
The computational model first calculated the $\Delta n$ and thickness of each cellulose lamina layer.
Along the light path, the direction of the fibrils changed layer by layer. Therefore, the total optical retardation of the entire cell wall was the vectorial summation of the retardation contributed by each layer of cellulose. The calculated OPD distributions were shown in Fig.~\ref{fig:Fig_Xylem_MFA_5modes_opt_cut_half}(b1-5).

With crossed P and A, the colors are darkness and gray.
The five structures were indistinguishable. Incorporating a retardation wave plate was essential to generate unique color for each of them.

With P-A-$\gamma =$ $0^\circ$-$90^\circ$-$45^\circ$, helix and twisted helix showed strong signal of fibrils in subtractive position, which was yellow, while the line, ring, and crossed helix were magenta (Fig.~\ref{fig:Fig_Xylem_MFA_5modes_opt_cut_half}(d1-5)).

With P-A-$\gamma =$ $45^\circ$-$135^\circ$-$90^\circ$, the line displayed blue (additive) and ring showed yellow (subtractive) (Fig.~\ref{fig:Fig_Xylem_MFA_5modes_opt_cut_half}(e1) and (e3)).

Twisted helix showed gradual color change from yellow/blue (inside) to magenta (outside) on the rectangular cross sections because of the twisting structure (Fig.~\ref{fig:Fig_Xylem_MFA_5modes_opt_cut_half} (d5) and (e5)).

Crossed fibrils canceled out the optical retardation from each layer, so the center curvy part of the sample appeared magenta regardless of the P-A-$\gamma$ settings (Fig.~\ref{fig:Fig_Xylem_MFA_5modes_opt_cut_half} (d4) and (e4)). This highly distinctive feature caused the crossed helix to stand out among all the cases.

The simulation confirmed that the birefringent-colored optical profiles of the longitudinal sections showed distinct characteristics of each cell wall structure. This represented a significant advancement in this research. 
Each fibril structure had its unique color distribution. By rotating the P-A-$\gamma$ and observing the color changes, the type of the cellulose organization could be clearly recognized.

\section{Experimental results}

\subsection{Verification of the birefringent-colored profiles}

\begin{figure*}
\centering
\includegraphics[width=1.0\textwidth]{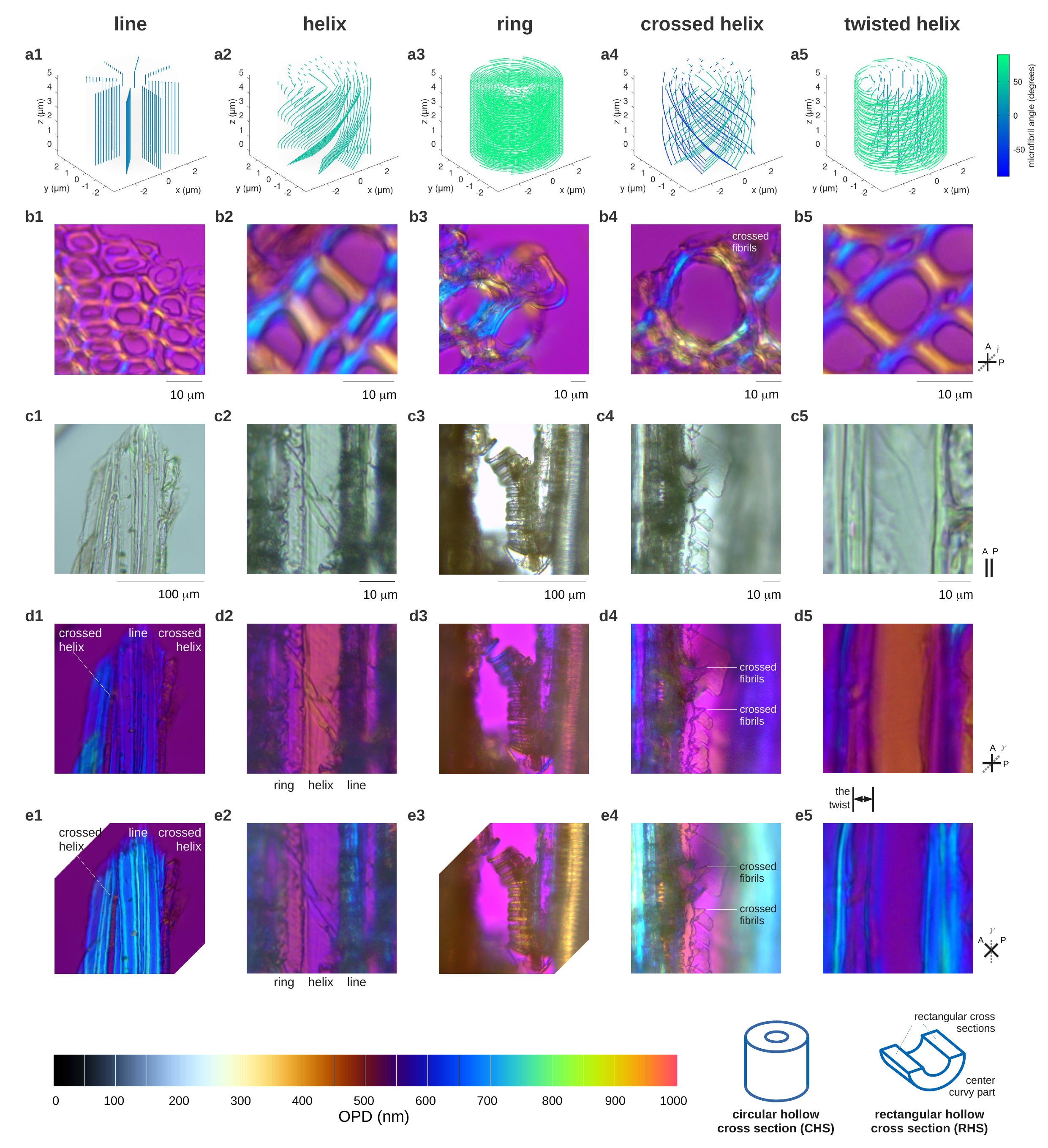}
\caption{POM images of xylem cells.
(a1-5) Schematics of the five cell types.
(b1-5) Transverse cross sections.
Longitudinal cross sections with
(c1-5) parallel A and P 
(d1-5) P-A-$\gamma$= $0^\circ$-$90^\circ$-$45^\circ$
(e1-5) P-A-$\gamma$= $45^\circ$-$135^\circ$-$90^\circ$.
P, polarizer.
A, analyzer.
$\gamma$, slow axis of the 530~nm retardation plate.
}
\label{fig:Xylem_exp}
\end{figure*}

Slices of \textit{Eucalyptus grandis} were placed under Zeiss Primotech.
The twig and trunk of the tree were sliced with a sharp blade.
The thickness of the specimen ranged from 5~\textmu m to 10~\textmu m.
The specimens were soaked in glycerol for optical clearing and preserving the microstructures.

The residual fibers at the cut, the cracks, and the surface of the cell walls revealed the fibril orientation.
Five cells were selected as representatives, because their fibril structures were clearly visible and confirmed to be line, helix, ring, crossed helix, and twisted helix.

The transverse cross sections were shown in Fig.~\ref{fig:Xylem_exp}(b1-5).
The line cell was mostly magenta.
The rest of the cases looked similar.
The twisted helix had gradual magenta (in) to blue/yellow (out) color distribution, which required careful inspection to discern.
In all fairness, relying solely on the transverse-sectional cuts could have easily led to misinterpretation.

Inspection of the longitudinal sections proceeded in three steps.
Firstly, the plain fiber texture and morphology showed the cell wall architectures directly (Fig.~\ref{fig:Xylem_exp}(c1-5)). The parallel P and A eliminated scattering light, improving the resolution.
Secondly, the P-A-$\gamma$ were applied to the exact same cell wall, and the textures were rendered with birefringent colors (Fig.~\ref{fig:Xylem_exp}(d1-5)).
Finally, the P-A-$\gamma$ were rotated, so the $\gamma$ aligned with the cell axes  (Fig.~\ref{fig:Xylem_exp}(e1-5)).
The colors presented by the actual samples and the colors predicted with simulation were consistent. The characteristic features were summarized in Table~\ref{table:ColorTags_5helix}.
In Fig.~\ref{fig:Fig_Xylem_MFA_5modes_opt_cut_half}(d1) and (e1), the bundle of lines (blue) and two crossed helices (magenta) were clearly visible.
Fig.~\ref{fig:Fig_Xylem_MFA_5modes_opt_cut_half}(d2) and (e2) included the line, helix, and ring cells, showing their distinct magenta, blue, and yellow colors when the P-A-$\gamma$ rotated.
Fig.~\ref{fig:Fig_Xylem_MFA_5modes_opt_cut_half} (d3) and (e3) clearly showed the fibril trend of a spring-like ring cell and the birefringent color.
The crossed fibrils (Fig.~\ref{fig:Fig_Xylem_MFA_5modes_opt_cut_half}(d4) and (e4)) and the twisted fibrils (Fig.~\ref{fig:Fig_Xylem_MFA_5modes_opt_cut_half}(d5) and (e5)) were marked in the pictures.

Using birefringent colors as tags to identify the cell wall structure yielded accurate results, as confirmed by the microscopic image of the fibril patterns.
This experiment demonstrated that birefringent colors could indicate the three-dimensional helical structure of the cell wall. The identification was easy, accurate, and reliable.

\begin{table*}
\caption{\label{table:ColorTags_5helix}Characteristic color distributions of the five typical cell wall structures.}
\begin{ruledtabular}
\begin{tabular}{cccccc}
\makecell[c]{}
&\begin{minipage}{0.12\textwidth}
      \includegraphics[width=\linewidth]{Xylem_3DPipe_line.png}
    \end{minipage}
&\begin{minipage}{0.12\textwidth}
      \includegraphics[width=\linewidth]{Xylem_3DPipe_helix.png}
    \end{minipage}
&\begin{minipage}{0.12\textwidth}
      \includegraphics[width=\linewidth]{Xylem_3DPipe_ring.png}
    \end{minipage}
&\begin{minipage}{0.12\textwidth}
      \includegraphics[width=\linewidth]{Xylem_3DPipe_cross.png}
    \end{minipage}
&\begin{minipage}{0.12\textwidth}
      \includegraphics[width=\linewidth]{Xylem_3DPipe_twist.png}
    \end{minipage} \\
\textbf{}
& \textbf{Line}
& \textbf{Helix}
& \textbf{Ring}
& \textbf{Crossed helix}
& \textbf{Twisted helix}\\
\hline
\makecell[c]{\textbf{Transverse cut in} \includegraphics[width=0.05\textwidth]{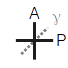}}
& \textcolor{magenta}{Magenta}
& Orthogonal cross
& Orthogonal cross
& Orthogonal cross
& \makecell[c]{\textcolor{magenta}{Magenta} (in) $\rightarrow$ \\ orthogonal cross (out)}\\
\hline
\makecell[c]{\textbf{Transverse cut in} \includegraphics[width=0.05\textwidth]{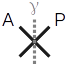}}
& \textcolor{magenta}{Magenta}
& Diagonal cross
& Diagonal cross
& Diagonal cross
& \makecell[c]{\textcolor{magenta}{Magenta} (in) $\rightarrow$ \\ diagonal cross (out)}\\
\hline
\makecell[c]{\textbf{Longitudinal cut in} \includegraphics[width=0.05\textwidth]{PAgamma_ortho.png}}
& \textcolor{magenta}{Magenta}
& \textcolor{Dandelion}{Yellow}
& \textcolor{magenta}{Magenta}
& \textcolor{magenta}{Magenta}
& \makecell[c]{ Center curvy part\\ \textcolor{Dandelion}{Yellow} \\ Rectangular walls \\ \textcolor{Dandelion}{yellow} (in) $\rightarrow$ \textcolor{magenta}{magenta} (out)}\\
\hline
\makecell[c]{\textbf{Longitudinal cut in} \includegraphics[width=0.05\textwidth]{PAgamma_dia.png}}
& \makecell[c]{\textcolor{blue}{Blue} \\ \textcolor{cyan}{Cyan}}
& \textcolor{magenta}{Magenta}
& \textcolor{Dandelion}{Yellow}
& \textcolor{magenta}{Magenta}
& \makecell[c]{ Center curvy part \\ \textcolor{magenta}{Magenta} \\ Rectangular walls \\ \textcolor{blue}{Blue} (in) $\rightarrow$ \textcolor{magenta}{magenta} (out)}\\
\end{tabular}
\end{ruledtabular}
\end{table*}

\subsection{Identifying cell wall structures in a composite of various cells}

\begin{figure*}
\centering
\includegraphics[width=1.0\textwidth]{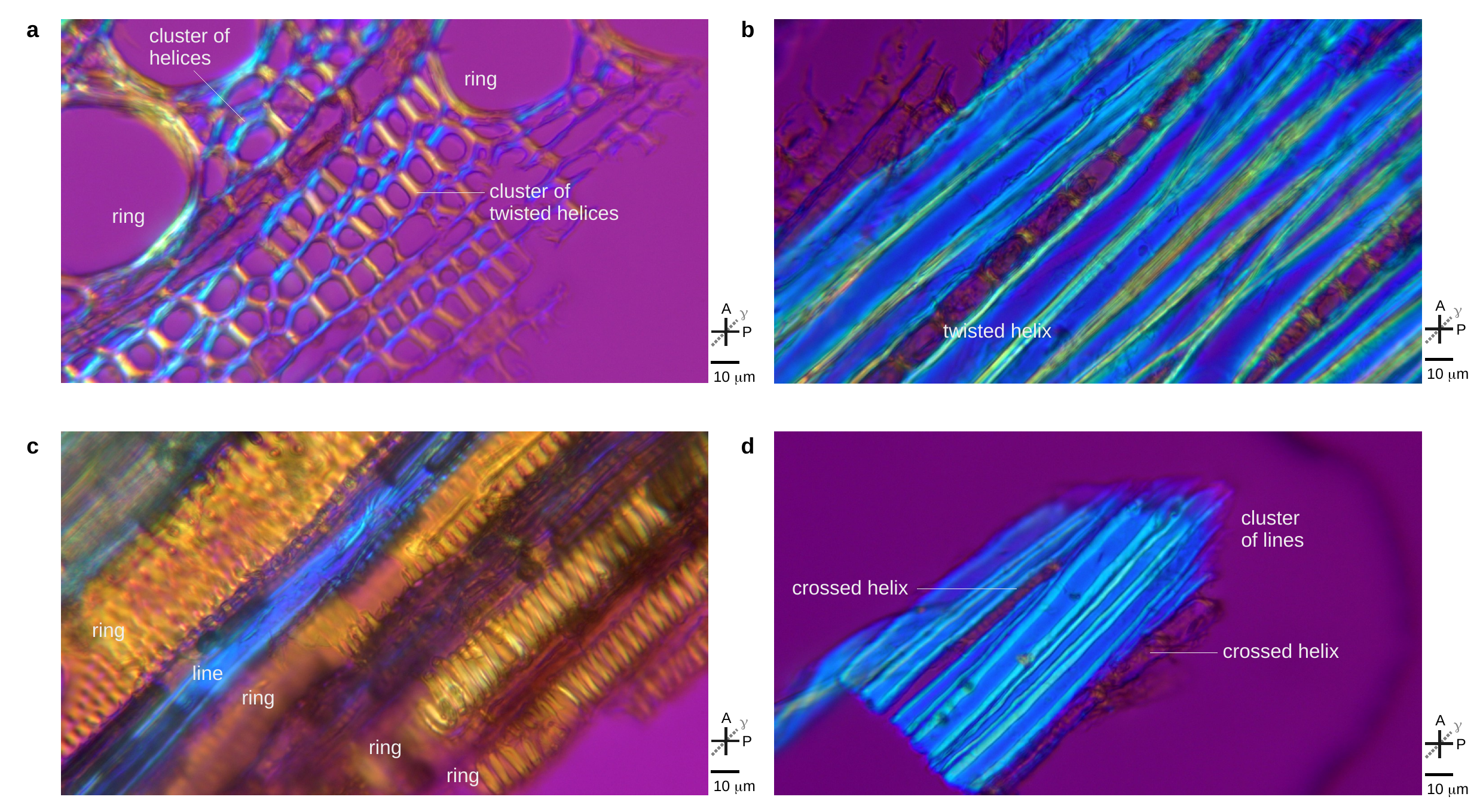}
\caption{Full-field POM images of xylem tissue.
(a) Transverse (b) longitudinal sections of trunk.
(c) Longitudinal cross section of twig.
(d) Bark tissue.
P, polarizer.
A, analyzer.
$\gamma$, slow axis of the 530~nm retardation plate.
}
\label{fig:Xylem_exp_full_field}
\end{figure*}

Four wood slices were taken from different parts of the eucalyptus tree (Fig.~\ref{fig:Xylem_exp_full_field}).
Each sample was a tissue composed of multiple types of cells.
The POM optical profiles clearly showed the five cell walls in different birefringent colors, even when the samples were complex compositions.

The trunk was composed of rings ($6\%$), helices ($27\%$) and twisted helices ($67\%$) (Fig.~\ref{fig:Xylem_exp_full_field}(a)).
The rings were large, elastic, and deformable for high speed water transport, while the helices could twist while maintaining the opening under the enormous hoop stress during water uptake.
The clusters of the twisted helices supported the soft, elastic vessel cells (Fig.~\ref{fig:Xylem_exp_full_field}(b)).
The soft foamy part at the center of the twig had a large portion of spring-like ring cells supported by the rigid line cells (Fig.~\ref{fig:Xylem_exp_full_field}(c)).
The bark of the twig was a bundle of line cells, which were rigid and easily fractured. Two out of ten cells in the bundle were crossed helical cells (Fig.~\ref{fig:Xylem_exp_full_field}(d)).
They looked similar in a regular micrograph (Fig.~\ref{fig:Fig_Xylem_MFA_5modes_opt_cut_half}(c1)). In the POM image, the lines were blue and crossed helices were magenta, which were clearly discernible.

Fig.~\ref{fig:Xylem_exp_full_field} revealed the irreplaceable advantage of the birefringent-colored labels.
Optical and electron microscopes were limited to display the shape and surface topography. When the cells had similar transparent texture and similar dielectric/refractive properties, it was difficult to identify the cells in a complex composite. 
The POM images overcame the obstacle with the brilliant blue, cyan, magenta, and yellow colors, rendering the various types of the cells clearly visible in one glance.

\section{Discussion}

\subsection{Advantages of polarized optical microscopy for wood analysis}

Polarized optical microscopy offers abundant information holistically in one picture. The sample can stay in water or glycerol, and the sample preparation does not destroy the 3D structure.
Electron microscopy is limited by the damaging sample preparation and the extremely localized view, and therefore statistical analysis is not feasible. Critically, electron microscopic pictures cannot distinguish the cellulose fibril from the lignin and hemicellulose matrix, causing lots of confusion and inconsistent results in the past decades.
Our results proved that POM was very sensitive to the birefringent cellulose, while the optically isotropic matrix did not interfere with the measurement.
Leveraging the interaction between polarized light and the cellulose effectively improved the 3D and statistical analysis of xylem cell walls.

\subsection{Sample preparation requirements for optimal results}

Dry wood has poor light transmittance and produces strong edge diffraction, resulting in dark and blurry images. Soaking wood slices in water or glycerol restores the transparency and the 3D microstructure. Glycerol, water, and cellulose have similar refractive indices, so the medium effectively eliminates the edge diffraction.

The sections need to be cut precisely along the axial axis or the transverse plane. A slanted cut introduces systematic shift of the OPD and the colors, which can lead to inaccurate identification.


Cell walls with high cellulose density (the trunk tissue) display saturated brilliant colors. On the other hand, the low-density ones (the twig tissue) look transparent. The background shows through the tissue, resulting in uninterpretable images.
Based on our practical experience so far, wood sections generally display clear birefringent colors. The cellulose density of leaves, epidermis, and grasses is insufficient to produce birefringent color that is discernible to the naked eye.

\subsection{Potential to AI-assisted automated analysis}

In POM images, dozens of cells and their colors can be observed and analyzed  simultaneously.
Therefore, we can count the number of each cell type, label their location, and determine the size of their clusters. The relationships between the cells are clearly visible. The birefringent-colored optical profile with an optical model of accurate interpretation enables statistical and spatial analysis.
Since the colored features of the five typical cell types are distinct, they are ideal for machine vision and machine learning programs to help process huge amounts of data. Neural Network algorithms should be able to recognize and summarize the optical profiles, enabling the fully automated analysis process.

\section{Conclusion}

Wood slices under a polarizing microscope display a spectrum of colors including cyan, blue, magenta, yellow, and bluish gray, rendering a beautiful image.
We delved into the physical mechanism that generated these colors and used the colors to tag the 3D structure of wood fibrils.
The optical model of the cellulose fibrils in the cell wall successfully demonstrated how light interacted with the birefringent cellulose, resulting in the interference colors.
The five typical cell wall architectures, line, helix, ring, crossed helix, and twisted helix, presented their signature colored textures.
Identifying the cell walls with morphology and with the birefringent-colored optical profiles led to the same results, proving that the optical indicators were efficient and accurate.
The colorful indicators are ideal for artificial-intelligence assisted algorithms, enabling the automated statistical and spatial analysis.
This method can support important research topics such as wood evolution and development, metamaterials with superior structural integrity inspired by wood, and wood-derived optical fibers, wave plates, and light diffusers for sustainable optoelectronics. This is also beneficial to the quality control of wood products and porous cellulose composites.
Most important of all, this research offers a unique perspective to read the information stored in wood. The five typical cell walls are like five letters of the tree's language. The polarized optical micrograph is the color-coded message that the tree sends to us through the light.  While appreciating the beauty of the colorful micrograph, we are reading the memories of a tree.

\appendix

\nocite{*}
\bibliography{aipsamp}

\end{document}